\documentclass[letterpaper,10pt]{article} 

\usepackage{opticameet3} 

\newcommand\authormark[1]{\textsuperscript{#1}}

\usepackage{amsmath,amssymb}
\usepackage[colorlinks=true,bookmarks=false,citecolor=blue,urlcolor=blue]{hyperref} 

\usepackage{graphicx}
\usepackage{amsmath}
\usepackage[dvipsnames]{xcolor}
\usepackage{hyperref}
\hypersetup{
    colorlinks=true,
    linkcolor=magenta,
    filecolor=blue, 
    urlcolor=blue,
    citecolor=blue,
    pdftitle={Overleaf Example},
    pdfpagemode=FullScreen,
    }
\usepackage{graphicx}
\usepackage{subcaption}
\usepackage{changes}

\definechangesauthor[name={Paolo}, color=teal]{pc}

\definechangesauthor[name={Elena}, color=blue]{er}

\begin{document}

\title{Temporal hopping dynamics in exciton-polariton condensation}

\author{Elena Rozas,\authormark{1,\dag,*} Wojciech Bukalski,\authormark{2,\dag}  Yannik Brune,\authormark{1} Adbhut Gupta,\authormark{4} Kirk Baldwin,\authormark{4} Loren N. Pfeiffer,\authormark{4} Hassan Alnatah,\authormark{5} Jonathan Beaumariage,\authormark{6} David W. Snoke,\authormark{6} Paolo Comaron,\authormark{3,2} Marzena H. Szyma\'{n}ska,\authormark{2} and Marc A{\ss}mann\authormark{1}}

\address{\authormark{1}Department of Physics, TU Dortmund University, 44227 Dortmund, Germany\\
\authormark{2}Department of Physics and Astronomy, University College London, WC1E 6BT London, UK\\
\authormark{3}CNR NANOTEC, Institute of Nanotechnology, 73100 Lecce, Italy\\
\authormark{4}Department of Electrical Engineering, Princeton University, Princeton, 08544 New Jersey, USA\\
\authormark{5}Joint Quantum Institute, University of Maryland, 4254 Stadium Dr., College Park, 20742 Maryland, USA\\
\authormark{6}Department of Physics, University of Pittsburgh, Pittsburgh, 15218 Pennsylvania, USA\\

\authormark{\dag}These authors contributed equally to this work.}

\email{\authormark{*}e-mail: elena.rozas@tu-dortmund.de}

\begin{abstract}
Polariton condensates provide a versatile platform for exploring non-equilibrium phase transitions and collective phenomena in open quantum systems. Near the condensation threshold, these systems are particularly sensitive to fluctuations and instabilities, which can strongly influence the condensate formation. Using optical trapping and homodyne detection, we directly access the photon statistics and second-order correlation function $g^{(2)}(0)$ of the condensate. We show that polariton condensation near the threshold is not a purely static transition, but instead undergoes a dynamical regime characterized by stochastic hopping between condensed and non-condensed states. These intermittent dynamics are accompanied by a gradual reduction of $g^{(2)}(0)$ towards unity, revealing the progressive build-up of coherence even in the presence of strong temporal fluctuations. Numerical simulations, based on a stochastic Truncated Wigner description of the driven–dissipative polariton field, reproduce these dynamics and capture the essential role of noise and reservoir interactions. This work demonstrates that the observed temporal hopping is an intrinsic feature of polariton condensation, providing a dynamical perspective that goes beyond static descriptions of the condensation phase transition.
\end{abstract}

\section*{Introduction}

Macroscopic coherence is a central feature of many collective phenomena in driven-dissipative systems, ranging from lasers to cold atoms \cite{Carusotto_2013, Walker2018, Tomita2017,Sharma2021}. Understanding how coherence emerges and evolves in time is crucial for fundamental studies of phase transitions and quantum technologies. Addressing such dynamics requires systems in which both interactions and dissipation can be directly accessed and controlled. In this context, exciton-polariton condensates emerge as powerful systems for exploring macroscopic quantum features in driven-dissipative environments. Owing to their hybrid light-matter nature, polaritons combine strong nonlinear interactions with direct optical access, enabling the study of non-equilibrium collective phenomena such as superfluidity \cite{Amo2009, Lerario2017}, quantized vortices \cite{Lagoudakis2008,Choi2022,Guerrero2025} and Bose-Einstein condensation \cite{Kasprzak2006,Balili2007,Dusel2020,Alnatah2025Jan}. A defining feature of polariton condensation is the emergence of coherence commonly revealed by the macroscopic occupation of a single mode followed by narrowing of its spectrum. In most experimental realizations, these properties are inferred from time-averaged observables, implicitly assuming that the condensed phase, once formed, is dynamically stable \cite{Ferrier2011,Galbiati2012,Byrnes2014,Betzold2020,Jin2025}. 

Advances in microcavity design have enabled polariton systems with exceptionally long lifetimes, substantially enhancing the condensate coherence by promoting efficient thermalization and reducing sensitivity to external noise sources \cite{Nelsen2013,Steger2013,Sun2017,Alnatah2024May,2025persistent,Brune2025}. While these features favor the formation of highly coherent condensates, extended lifetimes also grant access to slow intrinsic dynamics that are typically inaccessible in shorter-lived systems. A powerful route to access such dynamics is provided by engineered optical confinement. In  particular, ring-shaped optical traps have proven to efficiently separate the condensate from the incoherent reservoir, minimizing reservoir-induced dephasing and allowing polaritons to relax into the center of the trap \cite{Askitopoulos2013,Cristofolini2013,Orfanakis2021}. In the vicinity of the condensation threshold ($P_{th}$), driven-dissipative systems are expected to be especially susceptible to fluctuations and instabilities. In this regime, the condensate occupation remains low and the balance between gain and losses is fragile, allowing density fluctuations and mode competition to strongly influence the formation of the condensate \cite{Alnatah2024}. As a result, condensation cannot be regarded as a purely static transition. Therefore, accessing the temporal nature of these fluctuations requires time-resolved measurements of observables capable of probing the system on timescales faster than the intrinsic dynamics. Such measurements remain relatively scarce, despite their importance for understanding nonequilibrium phase transitions and the dynamical formation of coherent states in open quantum systems. 

In this work, we directly probe the temporal dynamics and stability of a polariton condensate by combining ring-shaped optical confinement with time-resolved homodyne detection. These dynamics are examined through intensity correlations and, in particular, via the second-order correlation function $g^{(2)}(0)$. We uncover pronounced temporal hopping dynamics in a narrow pump power range around the condensation threshold, characterized by the intermittent formation and disappearance of the condensate. With increasing excitation power, the hopping dynamics is suppressed, and the condensate enters a permanently stable phase. Numerical simulations based on a stochastic description of the polariton field within the Truncated Wigner Approximation reproduce these dynamics, and identify their origin in intrinsic driven-dissipative fluctuations near the threshold rather than extrinsic noise sources. Our results reveal that polariton condensation near the threshold proceeds through a dynamically unstable regime in which coherence is only intermittently established, providing a dynamical perspective on condensation beyond static threshold criteria.

\section*{Experimental platform}

\subsection*{Sample and setup}
The experiments were performed in a high-Q ($\simeq$ 300 000) $3\lambda /2$ GaAs microcavity. The structure incorporates 12 GaAs quantum wells, with 7 nm of thickness each. The QWs are arranged in three sets of four, each positioned at an antinode of the cavity field. The cavity is embedded between two distributed Bragg reflectors composed of 32 (top) and 40 (bottom) $\text{AlAs}/\text{Al}_{0.2}\text{Ga}_{0.8}\text{As}$ pairs, yielding a long polariton lifetime of approximately 300 ps. All measurements were conducted in a region where the cavity-exciton detuning was $\delta_{C-X}=0.4$ meV, estimated from the sample characterization in Ref. \cite{Beaumariage2024}. To decouple the condensate from the fluctuating reservoir, an annular optical trap was created by shaping a non-resonant cw laser, at 1.746 eV, with a spatial light modulator (SLM), into a ring-shaped excitation profile with a diameter of 9.6 µm on the sample surface. A microscope objective of $NA=0.26$ was used for both excitation and collection of the PL. The collected PL was polarization filtered along the cavity main axis, which corresponds to the polarization of the condensate at high pump powers. More details can be found in Ref. \cite{Brune2025}. This geometry confines the condensate to the trap center, suppressing the decoherence effects from carrier-polariton interactions and enabling a condensate close to thermal equilibrium.

\subsection*{Condensation and instability window}

\begin{figure}[h]
            \centering
            \includegraphics[width=\textwidth]{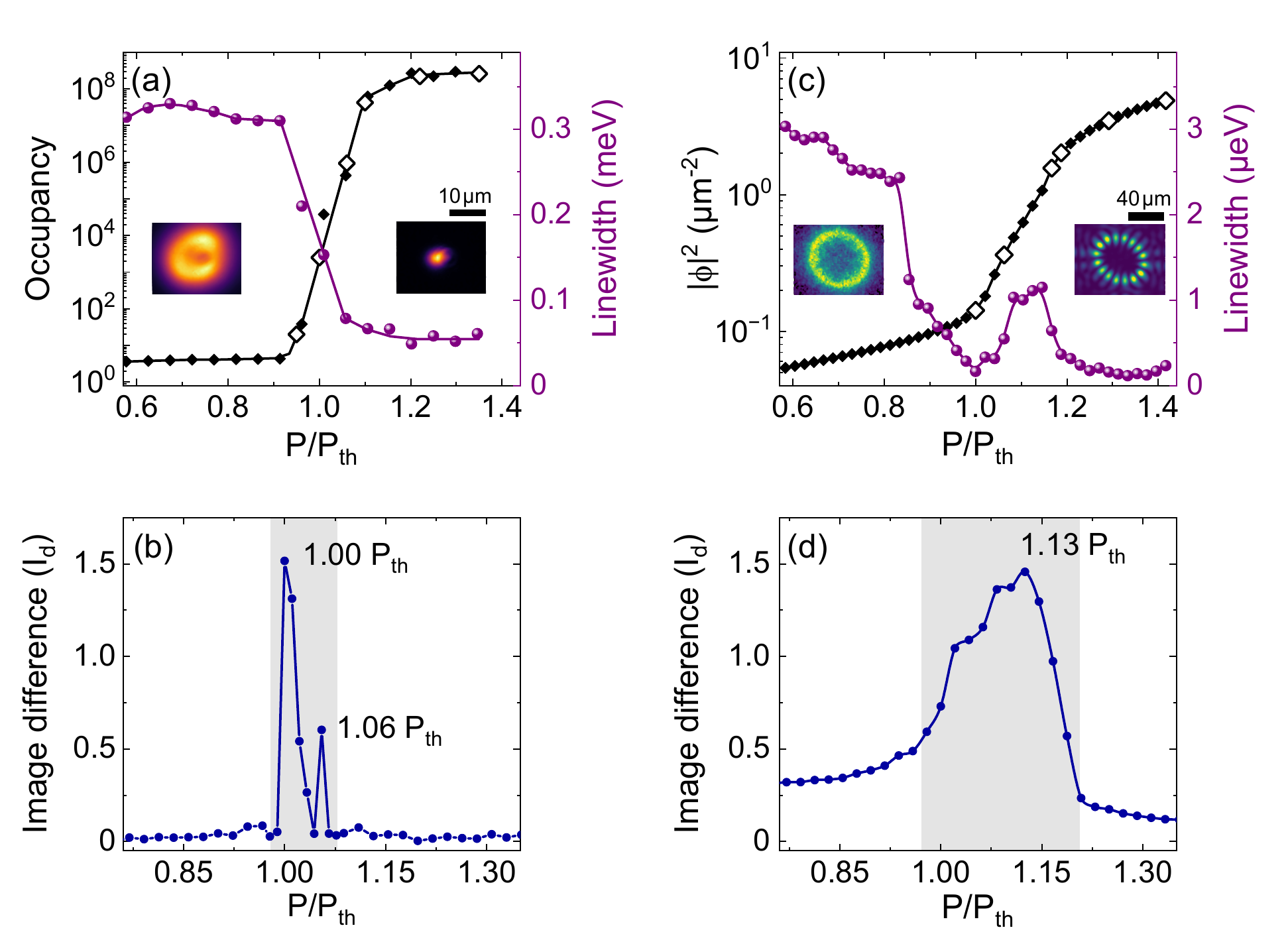}
            \caption{\textbf{Threshold behavior and instability of the polariton condensate.} (a) Measured polariton occupancy (black) and its corresponding PL linewidth (purple) as a function of the normalized pump power. The solid lines serve as guides to the eye. Empty diamonds highlight the pump powers at which detailed measurements are performed: $0.95 \, P_{th}, 1.00 \, P_{th}, 1.06 \, P_{th}, 1.10 \, P_{th}, 1.22 \, P_{th}$ and $1.35 \, P_{th}$. Two insets show the real-space image of the uncondensed polaritons (left) and the trapped polariton condensate (right). (b) Average image difference $I_{d}$, as defined in Eq. (\ref{eq:Id}), around the threshold area. The gray shaded area indicates the range of powers where instabilities of the polariton condensate are observed. (c) Numerical simulation equivalent of (a), showing the average density of the polariton field (black) as well as the evolution of the linewidth (purple). Here, the simulations at pumping powers $1.00 \, P_{th}, 1.06 \, P_{th}, 1.17 \, P_{th}, 1.19 \, P_{th}, 1.29 \, P_{th}$ and $1.42 \, P_{th}$ were selected for further analysis. (d) Calculated average image difference from the numerical simulations. In (c) and (d), the pumping threshold corresponds to the mean field value, obtained through Eqs. (\ref{polariton_condensate_eqn}-\ref{reservoir_eqn}) without the inclusion of $dW_{c}$.}
            \label{fig:Fig1}
        \end{figure}

The condensation dynamics in a ring-shaped trap is governed by the interplay between polariton relaxation, spatial confinement, and the driven-dissipative nature of the system. By increasing the excitation power these dynamics are triggered, resulting in a polariton system that evolves from a dilute incoherent population towards a macroscopically occupied coherent state. This transition is directly reflected in both the spatial distribution and the spectral emission, which provide clear signatures of the condensation process and its threshold, as illustrated in Fig. \ref{fig:Fig1}. Below $P_{th}$, the polariton occupancy remains low, with an emission localized on the ring-shaped potential, as shown in the left inset of Fig. \ref{fig:Fig1}(a). In this regime, the emission displays a generally broad linewidth, and no condensate is formed yet. Near $P_{th}$, the polariton population builds up in the center of the trap, leading to a rapid nonlinear increase in occupancy, spanning nearly eight orders of magnitude. This transition is accompanied by a pronounced linewidth narrowing, concurrent with the emergence of coherence. Above threshold, the growth in occupancy saturates, and a well-defined condensate forms inside the trap. This final state is shown in the right inset of panel (a).

Although these spectral properties provide clear signatures of the condensation process, they offer only limited insight into the dynamical behavior of the system. In particular, the smooth increase in occupancy can conceal temporal fluctuations and instabilities when averaged over timescales longer than the intrinsic dynamics of the condensate. To provide an initial indication of these effects, we introduce a quantity sensitive to shot-to-shot acquisitions by defining the normalized average image difference $I_{d}$ between five consecutive real-space emission frames, each integrated over 100 ms. This quantity provides a sensitive measurement of temporal changes in intensity, and is defined as \cite{Alnatah2024},
  \begin{align}
    I_d = \frac{\left\langle RMS\left( I_{n+1} - I_n \right) \right\rangle}{\left\langle I_n \right\rangle} \text{,}
    \label{eq:Id}
    \end{align}
where RMS denotes the root mean square. For each pump power, $I_d$ is normalized by the mean emission intensity for that specific power to allow for direct comparison across all sets. The evolution of $I_d$ along different pump powers is shown in Fig. \ref{fig:Fig1}(b). Below $P_{th}$, $I_d$ remains close to zero, indicating that consecutive real-space images do not exhibit significant temporal variations. As the excitation power approaches the threshold, $I_d$ develops strong fluctuations, triggered by the onset of dynamical instabilities. This unstable region, highlighted by a gray area, extends up to approximately $1.08\, P_{th}$. Beyond this point, the suppression of temporal fluctuations renders the condensate stable and $I_d$ decreases again to zero.

\section*{Theoretical framework}

Let us now focus on the theoretical description of polariton condensation. Within the Truncated Wigner Approximation (TWA), the elliptically pumped system can be described by a stochastic partial differential equation for the condensate at the lower polariton branch coupled to a rate equation describing the excitonic reservoir that feeds the condensate \cite{Alnatah2024, Carusotto_2013, Wouters_2009}:

\begin{equation}
\label{polariton_condensate_eqn}
    i \hslash d\psi(\mathbf{r}) = dt \Bigg[ (i\beta - 1)\frac{\hslash^2 \nabla^2}{2m} + g_{\mathrm{c}} |\psi(\mathbf{r})|^{2} + g_{\mathrm{R}} n_{\mathrm{R}}(\mathbf{r}) + \frac{i\hslash}{2}\left(R n_{\mathrm{R}}(\mathbf{r}) - \gamma_{\mathrm{c}} \right) \Bigg]\psi(\mathbf{r}) + i \hslash dW_c(\mathbf{r}, t),
\end{equation}

\begin{equation}
\label{reservoir_eqn}
    \frac{d}{dt} n_{\mathrm{R}}(\mathbf{r}) = P_r(\mathbf{r}) - \left( \gamma_{\mathrm{R}} + R|\psi(\mathbf{r})|^2 \right) n_{\mathrm{R}}(\mathbf{r}),
\end{equation}

where the complex Gaussian stochastic noise term, $dW_c(\mathbf{r}, t)$, satisfies the correlations

\begin{equation}
    \langle dW_c(\mathbf{r}, t) dW_c(\mathbf{r'}, t) \rangle = 0,
\end{equation}

\begin{equation}
    \langle dW_c(\mathbf{r}, t) dW_c^*(\mathbf{r'}, t) \rangle = \frac{Rn_{\mathrm{R}}(\textbf{r}) + \gamma_{\mathrm{c}}}{2dV}\delta_{\mathbf{r}, \mathbf{r'}}dt.
\end{equation}

As such, the diffusive noise term accounts for quantum fluctuations introduced through coupling to the environment \cite{ Bobrovska_2015,comaron2020BKT}. Throughout Eqs. (\ref{polariton_condensate_eqn}-\ref{reservoir_eqn}), $\psi(\mathbf{r}, t)$ denotes the polariton field, while $n_{\mathrm{R}}(\mathbf{r})$ represents the density of the exciton reservoir. The nonrenormalised quantity $|\psi(\mathbf{r})|^2$ is used wherever the polariton density appears in these equations. Within the TWA description, this corresponds to omitting the commutator correction arising from symmetric operator ordering \cite{Wouters_2009, Olsen_2009}. Benchmarking indicates that this approximation does not qualitatively alter the dynamics, and the resulting small quantitative shift in the condensation threshold and average density facilitates further analysis (See Supplementary material). The effective polariton mass, $m$, enters through the kinetic term of Eq. (\ref{polariton_condensate_eqn}), which also introduces the phenomenological energy relaxation parameter $\beta$ \cite{Wouters_2010_1, Wouters_2010_2, Wouters_2012, Gladilin_2020}. The polariton-polariton and polariton-exciton interaction strengths are denoted by $g_{\mathrm{c}}$ and $g_{\mathrm{R}}$, respectively. They are estimated as $g_{\mathrm{c}} = g_{\mathrm{ex}}|X|^4$ and $g_{\mathrm{R}} = g_{\mathrm{ex}}|X|^2$, where $g_{\mathrm{ex}}$ is the exciton-exction interaction strength and $|X|^2$ is the excitonic fraction determined by the Hopfield coefficient $X$ \cite{Estrecho_2018}. The excitonic fraction also determines the decay rate of the condensate, $\gamma_{\mathrm{c}}$, estimated as $(1 - |X|^2)/\tau_{\mathrm{ph}}$, where $\tau_{\mathrm{ph}}$ is the photon lifetime in the microcavity \cite{Estrecho_2018,Comaron_2025}. The stimulated scattering rate, $R = R_{0}g_{\mathrm{c}}/g_{\mathrm{R}}$, governs the transfer of excitons from the reservoir to the polariton condensate. This parameter duly presents itself as a source of loss in Eq. (\ref{reservoir_eqn}), with the remaining exciton loss originating from $\gamma_{\mathrm{R}}$, which represents their intrinsic decay rate. Pumping is incorporated through $P_r(\mathbf{r})$, which models the experimentally relevant elliptical pump geometry via $P_r(\mathbf{r}) = (n_0\gamma_{\mathrm{c}}/R)\exp[-\left(\mathcal{E}/2\sigma^2\right)^2]$, where $n_0$ represents the initial density of the exciton reservoir, and where $\sigma$ determines the effective width of the pumping ring. The equation for an ellipse, tilted through an angle $\theta$, is given by $\mathcal{E} = Ax^2 + Cy^2 + Bxy - 1$, with $A = (\cos^2\theta)/a^2 + (\sin^2\theta)/b^2$, $B = 2\sin\theta\cos\theta$, and $C = (\sin^2\theta)/a^2 + (\cos^2\theta)/b^2$, and as such $a$ and $b$ are the major and minor axes, respectively. The geometric parameters $b = 30 \ \mu\mathrm{m}$, $a = 1.1b$, $\theta = 0.5 \ \mathrm{rad}$, and $\sigma = 1.53 \ \mu\mathrm{m}$ reproduce the experimental configuration on a $128 \times 128$ grid with lattice spacing $1.17 \ \mu\mathrm{m}$, albeit with a radius six times larger than in the experiment. These values were adopted from \cite{Alnatah2024}, together with the remaining parameters: $\beta = 0.02$, $m = 4.2 \times 10^{-5} m_e$, where $m_e$ is the electron mass, $g_{\mathrm{ex}} = 2 \ \mu\mathrm{eV}\mu\mathrm{m}^2$,  $R_0 = 2 \times 10^{-3} \ \mu\mathrm{m}^2\mathrm{ps}^{-1}$, $\tau_{\mathrm{ph}} = 135 \ \mathrm{ps}$, $|X|^{2} = 0.1$, and $\gamma_{\mathrm{R}} = 10^{-3} \ \mathrm{ps}^{-1}$. The larger system size and the more photonic character of the polaritons are motivated in the Supplementary material.

In addition to the dynamical noise introduced through $dW_c$, the simulations require an additional source of initial noise. This inclusion is essential to differentiate the single-time dynamics between realizations, determining the onset time, duration, and mode-selectivity of hopping. For the parameter regime considered here, a total evolution time of 500 ns was found to be sufficient for the system to relax and for the subsequent hopping dynamics to emerge. With the theoretical framework now established, we turn to the numerical simulations presented in Fig. \ref{fig:Fig1}(c), which successfully reproduce the main features observed experimentally. In the low power regime, the system is dominated by an incoherent population whose density remains largely confined to the pumping ring, as illustrated in the left inset of panel (c). As the condensation threshold approaches, the model captures the onset of a coherent population localized within the trap. Here, the condensate adopts a sixteen-lobe whispering-gallery spatial configuration, shown in the right inset of panel (c), reflecting the larger trap dimensions used in the theoretical model \cite{Sun_2018}.

To further quantify temporal fluctuations, Fig. \ref{fig:Fig1}(d) shows the normalized average image difference extracted from the simulations. The numerical model reveals the presence of an unstable region between $0.95\, P_{th}$ and $1.20\, P_{th}$. This region is characterized by strong mode competition and coincides with the minor linewidth broadening shown in Fig. \ref{fig:Fig1}(c), a feature absent in the experimental data presented in panel (b). This discrepancy arises because competing modes in the simulations are spaced more closely in energy than in the experiment, leading to a noticeable spectral broadening. In addition, the choice of integration domains in the numerical analysis, as well as the image integration settings, which differ from those used experimentally, may further contribute to the observed broadening. Details of the numerical procedure are provided in the Supplementary Material.

\section*{Temporal hopping dynamics}

Understanding how the condensate builds up and stabilizes requires access to its dynamics on short timescales. To achieve this with high temporal sensitivity, we employ a homodyne detection scheme, which allows us to access the intrinsic dynamics in the steady-state regime \cite{McAlister1997,Roumpos2013,Luders_2021}. In this configuration, a single-channel balanced detector is used to measure one field quadrature of the emitted signal by interfering it with a stable resonant local oscillator. Following the procedure described in ref.\cite{Brune2025}, data subsets of 30.000 quadrature samples are computed to obtain a temporal window of approximately 40 $\mu s$. This approach provides direct access to the temporal evolution of the photon number distribution and enables the reconstruction of the corresponding second-order correlation function $g^{(2)}(0)$, thereby allowing us to probe the emergence of coherence and the associated phase-driven fluctuations in the system \cite{Luders2018}.

For TWA simulations on a discretized spatial grid, the equal-time second-order correlation function is given by 

\begin{equation}
    g^{(2)}(\tau = 0) = \frac{\langle \psi^*\psi^*\psi\psi \rangle_W - \frac{2\langle \psi^*\psi \rangle_W}{dV} +\frac{1}{2dV^2}}{(\langle \psi^*\psi \rangle_W - \frac{1}{2dV})^2},
\end{equation}
where $\left< \cdots \right> _W$ denotes the stochastic Wigner average. Here, the Wigner correction terms introduce several pathologies for the observable. In particular, the numerator may transiently become negative, while the overall expression approaches unity asymptotically from below as we achieve condensation. Most critically, the denominator introduces a divergence when the mean density approaches $1/2dV$. For the parameter settings employed here, this corresponds to a physical density approximately equal to $0.36 \ \mu\mathrm{m}^{-2}$. In spatial regions where the local density is close to this value, the resulting $g^{(2)}(0)$ becomes nonphysical and can grow arbitrarily large. Since the system must inevitably pass through this density during the hopping process, the observable briefly loses reliability, necessitating particular care when interpreting equal-time correlation measurements. The impact of these corrections is discussed further in the Supplementary material. 

In direct analogy to the experiment, we track the temporal evolution of a single mode. Numerically, this is achieved by filtering $\psi$ in frequency space prior to the evaluation of TWA observables. The resulting single-mode dynamics is faithfully characterized by $g^{(2)}(0)$, provided that the selected mode rapidly acquires an occupation exceeding $0.36 \ \mu\mathrm{m}^{-2}$. Although the ground state is the first to gain a significant population upon crossing the condensation threshold \cite{Sun_2018}, its average occupation subsequently fails to surpass $0.36 \ \mu\mathrm{m}^{-2}$ for the pumping powers considered. We accordingly base our analysis on the first excited state instead, which becomes highly occupied and dominates the dynamics in the hopping region, ultimately outcompeting other modes to produce the spatial density profile shown in the inset of Fig. \ref{fig:Fig1} \cite{Alnatah2024}.

Resolving intermittent transitions requires working at the level of single shots, and as such, individual TWA realizations. This requirement poses another challenge for the TWA formalism, which typically attains observables by averaging over many stochastic realizations. Instead, we adopt a time-averaging strategy that mirrors the experimental procedure. More specifically, the field $\psi$ is sampled densely in time, 25000 samples in the present work, and the resulting data are averaged by using a moving temporal window. The relatively large window size of 500 samples employed in our case suppresses extremely short-lived excursions, preferentially revealing events associated with definite sustained disappearance. Each individual time sample is rendered relatively well-behaved by this averaging procedure. Nevertheless, additional filtering remains essential to mitigate contamination from densities close to the pathological value $0.36 \ \mu\mathrm{m}^{-2}$. To this end, $g^{(2)}(0)$ is first computed locally, in a piecewise fashion over the spatial grid at each time sample, enabling the straightforward identification of problematic spatial points. A filter of the form $0.5 \leq g^{(2)}(0) \leq 2.5$ is then applied to remove nonphysical divergences while accounting for the asymptotic behavior of the observable. An additional mask is subsequently applied to retain only those grid points belonging to the spatial profile of the tracked mode. The surviving points, numbering in excess of $\approx 10^2$ at minimum, are then averaged independently for each time sample. The resulting time series of averaged operator products enables a robust evaluation of $g^{(2)}(0)$ within the constraints discussed above.

\subsection*{Photon number}

\begin{figure}[t]
            \centering
            \includegraphics[width=0.8\columnwidth]{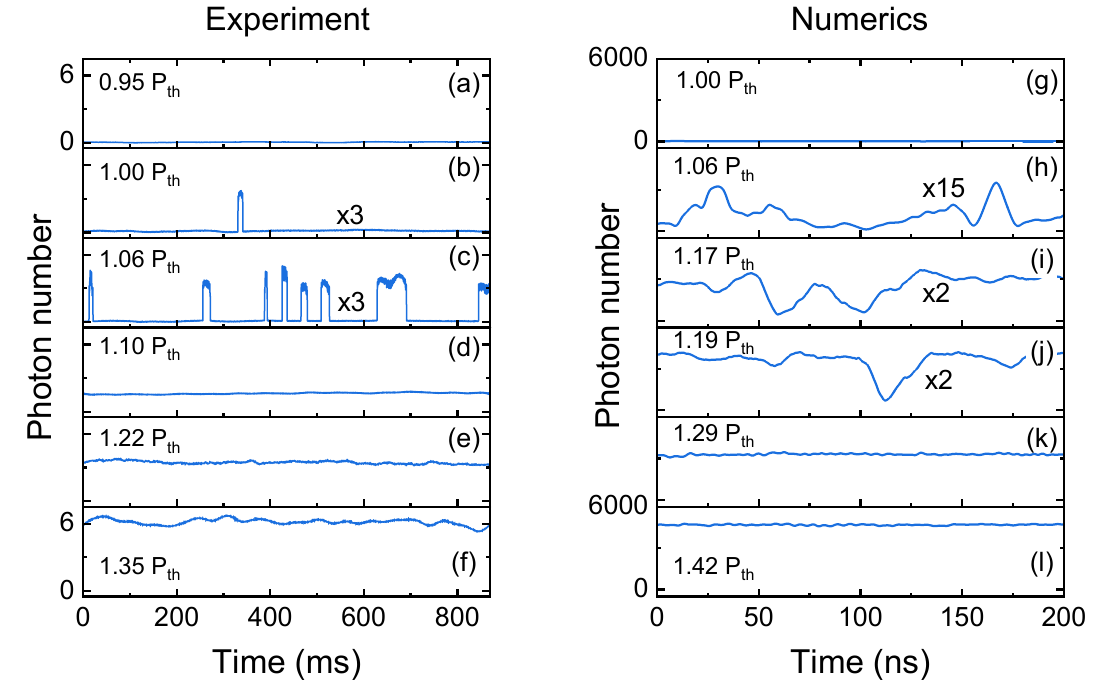}
            \caption{\textbf{Time-resolved photon number} Experimental (a-f) and numerically simulated (g-l) photon numbers with increasing pump power. Below threshold (a), no measurable signal is detected due to the strong blueshift of the emission, which prevents the spectral overlap with the local oscillator. Panels (b) and (c) are magnified by a factor of three for clarity.}
            \label{fig:Fig2}
        \end{figure}

Fig. \ref{fig:Fig2} presents the time-resolved photon number obtained from both experiment and numerical simulations for increasing pump power. Panels (a–f) show the experimental measurements, while panels (g–l) display the corresponding results from the theoretical model. To track the evolution of the polariton system across the different regimes, six representative pump powers around the condensation threshold are selected, spanning from the linear regime to the fully stable condensed phase. These powers are indicated by open diamonds in Fig. \ref{fig:Fig1}(a) and (c), respectively. Here, the time axis corresponds to the acquisition window of each measurement, with its origin defined by the start of the acquisition. Thus, it does not represent an absolute timescale of the condensate dynamics.

In the experiment, at $0.95\, P_{th}$, no emission is detected by the homodyne setup. This is primarily due to the large blueshift of $\simeq 5$ meV between the uncondensed polaritons and the condensate, which prevents the spectral overlap between the signal and the local oscillator. Note that the LO is adjusted to match only the condensate energy. Consequently, the measured signal is indistinguishable from vacuum noise. At $1.00 \, P_{th}$, panel (b), the emergence of the condensate emission becomes detectable. A distinct peak appears for a short duration of 10 ms, marking the initial formation of the condensate at the center of the trap. A further increase of the pump power leads to clear hopping dynamics in the photon number, characterized by irregular, non-periodic fluctuations, reflecting the intermittent switching between states of low and high occupation. These fluctuations can be understood as the result of mode competition and stochastic gain-loss fluctuations near threshold, where small perturbations are sufficient to destabilize the condensate. This behavior persists up to $1.10\, P_{th}$, beyond which the unstable regime is overcome, the hopping dynamics vanish and the condensate acquires temporal stability with a photon number settling around 1.7 photons. Note that the quoted photon number corresponds to the detected photons in the homodyne scheme and thus, represents only a fraction of the cavity population. The actual polariton number is estimated to be $10^3$, placing the system well within the semiclassical regime. At higher powers, the emission intensity gradually increases reaching 6 photons at $1.35\, P_{th}$. The overall behavior observed here corresponds closely to the unstable region previously identified in Fig. \ref{fig:Fig1}(b).   

The polariton number fluctuations observed across the range of pump powers probed in the numerical simulations, Fig. \ref{fig:Fig2}(g-l), corroborate the experimental findings. In particular, the simulations confirm the non-periodic and transient character of the phenomenon, which is evident on both sides of the unstable region highlighted in Fig. \ref{fig:Fig1}(d). These fluctuations become increasingly frequent as we probe deeper into the hopping regime, approaching the peak observed in the image difference. Following this process in reverse, the fluctuations become progressively sparser in time as the pump power increases from $1.17\, P_{th}$ to $1.19\, P_{th}$, beyond which the condensate stabilizes and the fluctuations abruptly stop. Further increases in pump power only lead to an increase in the total polariton population, consistent with the experiment. We note that the total polariton numbers exceed those measured experimentally, owing to the more photonic character of the simulated system, as well as the larger analyzed spatial region.

\subsection*{Second-order correlation function}

\begin{figure}[t]
            \centering
            \includegraphics[width=0.8\columnwidth]{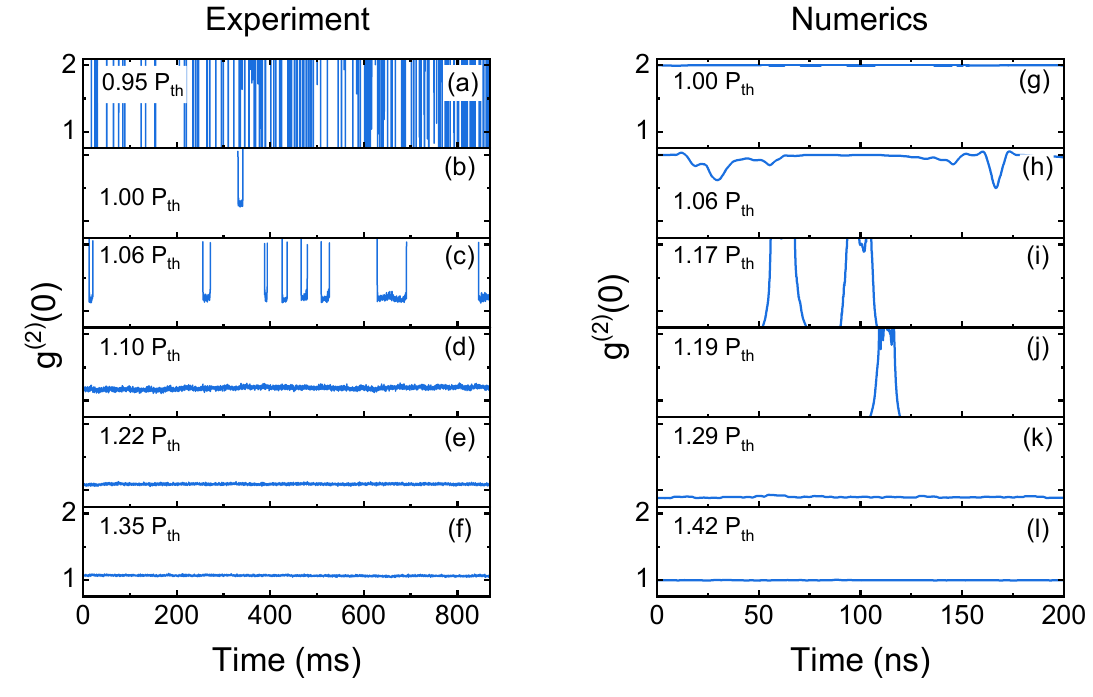}
            \caption{\textbf{Time-resolved second-order correlation function $g^{(2)}(0)$.} Experimental (a-f) and numerical (g-l) $g^{(2)}(0)$ with increasing pump power. Due to the mismatch between signal and LO at 0.95 P$_{th}$ (a), no signal is detected and therefore, $g^{(2)}(0)$ results in random values.}
            \label{fig:Fig3}
\end{figure}

The corresponding evolution of the second-order correlation function $g^{(2)}(0)$ for the same set of pump powers is presented in Fig. \ref{fig:Fig3} (a–f) and (g-l) for the experimental results and the theoretical model, respectively. As discussed in the photon number analysis, at a low excitation power of $0.95\, P_{th}$ in the experiment, the detected signal is indistinguishable from vacuum fluctuations. Therefore, the extracted $g^{(2)}(0)$ values are random and do not carry any physical meaning. These random values are omitted from the rest of the plots for clarity. Once the condensate becomes visible, see panel (b), $g^{(2)}(0)$ exhibits a sudden reduction to 1.28, signaling the onset of coherence in the polariton system. As the power is increased to $1.06\, P_{th}$, the system enters the irregular hopping regime previously identified in Fig. \ref{fig:Fig2}. In this regime, pronounced fluctuations coexist with a continued reduction of $g^{(2)}(0)$ down to 1.19, revealing a progressive build up of coherence despite the underlying dynamical instabilities. This behavior reveals that the establishment of coherence is not contingent on a steady-state condensate, but can instead emerge under strongly nonstationary conditions. Notably, the progressive reduction of $g^{(2)}(0)$ in the hopping regime demonstrates that coherence can develop independently of temporal stability, challenging the conventional association of condensation with stationary coherence. Beyond $1.10\, P_{th}$, the system finally transitions into a stable regime, where the hopping dynamics vanish and $g^{(2)}(0)$ saturates, indicating that the coherence is no longer limited by temporal hopping. Upon further increasing the power, coherence continues to improve, with $g^{(2)}(0)$ reaching a value of 1.06, consistent with a highly coherent condensate. 

The numerical simulations successfully reproduce this behavior and provide further insight into the underlying regimes. At densities an order of magnitude below the problematic value $1/2dV$, the system resides in a low-occupation regime dominated by Gaussian fluctuations generated by stochastic noise. In this limit, the equal-time observable is well-defined, yielding $g^{(2)}(0) \approx 2$, as expected for a thermal state. This behavior is exemplified at $1.00\, P_{th}$ in Fig. \ref{fig:Fig3}(g). At intermediate pump powers, corresponding to the unstable region highlighted previously, the polariton system reveals hopping dynamics that manifest as intermittent transitions between the well-defined thermal and condensed regimes, with such events becoming increasingly frequent within this region. Here, the tracked mode may temporarily vanish in favor of competing modes, or briefly lose any identifiable condensed mode altogether, reverting to a fully thermal-like state. During the hopping process, the system necessarily explores density ranges in which the equal-time observable becomes ill-defined. Nevertheless, provided the system remains predominantly condensed over the probed time window, the methodology outlined above remains applicable and enables hopping events to be identified. Although this prevents a faithful characterization at the peak of the hopping region, namely at $1.13\, P_{th}$, the behavior away from this point is sufficient to confirm the correspondence between experiment and model. Finally, at high densities well above $1/2dV$, the system crosses into a high-occupation regime governed primarily by classical drift. In this regime, the Wigner correction terms become negligible, and $g^{(2)}(0)$ approaches 1, consistent with a coherent condensed state with suppressed intensity fluctuations, as illustrated at $1.42\, P_{th}$.

\section*{Conclusions}

We have investigated the temporal dynamics of an optically confined exciton-polariton condensate by combining optical trapping with time-resolved homodyne detection under cw excitation. This approach provides direct access to the photon statistics and $g^{(2)}(0)$ function with high temporal resolution, enabling the characterization of condensation dynamics beyond conventional time-integrated observables. Our results demonstrate that polariton condensation near the threshold cannot be understood as a purely static transition. Instead, the system exhibits a pronounced dynamical regime in which coherence is only sporadically established. This regime is characterized by stochastic hopping dynamics, where the condensate undergoes irregular transitions between low and high occupation states. Importantly, these hopping events are accompanied by a progressive reduction of $g^{(2)}(0)$ towards unity, revealing the gradual build up of coherence despite the presence of strong temporal fluctuations.

By comparing experimental observations with numerical simulations based on TWA, we show that these fluctuations arise from the intrinsic driven-dissipative nature of the system. The theoretical framework explicitly accounts for noise originating from coupling to the environment and interactions with the incoherent reservoir, reproducing both the fluctuating dynamics and the transition into a stable condensed phase. This agreement demonstrates that the observed dynamics are not governed by extrinsic perturbations, but rather emerge from the fundamental stochastic processes inherent to polariton condensation. The identification of this transition from hopping to a stable phase provides a unified picture linking photon statistics, coherence, and stability across the condensation threshold region. These findings establish a dynamical perspective on condensation in non-equilibrium systems, highlighting the importance of temporal fluctuations in macroscopic coherence. Furthermore, they open the way to controlling and exploiting threshold phenomena in quantum-based technologies, where the interplay between noise, nonlinearity, and dissipation can be harnessed to engineer novel dynamical states. 

\section*{Funding}
The work conducted by the Dortmund group was supported by the QuantERA II Programme, funded by the EU H2020 research and innovation programme, GA No. 101017733 and by the DFG Grant No. 532767301. 
W.B. acknowledges support from the Engineering and Physical Sciences Research Council (EP/T517793/1 and EP/W524335/1).
The Princeton University portion of this research is funded in part by the Gordon and Betty Moore Foundation’s EPiQS Initiative, Grant GBMF9615.01 to Loren Pfeiffer.
P.C acknowledges support from “Quantum Optical Networks based on Exciton-polaritons” (Q-ONE, N. 101115575, HORIZON-EIC-2022-PATHFINDER CHALLENGES EU project), "Neuromorphic Polariton Accelerator" (PolArt, N.101130304, Horizon-EIC-2023-Pathfinder Open EU project), “National Quantum Science and Technology Institute” (NQSTI, N. PE0000023, PNRR MUR project), “Integrated Infrastructure Initiative in Photonic and Quantum Sciences” (I-PHOQS, N. IR0000016, PNRR MUR project).
M.S. acknowledges support from the Engineering and Physical Sciences Research Council Grant. No. EP/V026496/1.
Views and opinions expressed are, however, those of the author(s) only and do not necessarily reflect those of the European Union or European Innovation Council and SMEs Executive Agency (EISMEA). Neither the European Union nor the granting authority can be held responsible for them.

\section*{Acknowledgments}
We would like to thank D. Ballarini, A. Ferrier, F. Laussy, and D. Sanvitto for fruitful discussions. This research was partly carried out using the high performance computing cluster at the London Centre for Nanotechnology, and made use of the UCL Myriad High Performance Computing Facility (Myriad@UCL) and its associated support services.

\section*{Disclosures}
The authors declare no conflicts of interest.

\section*{Data Availability Statement}
Data underlying the results presented in this paper are available in Ref. \cite{mydata}.

\section*{Supplemental document}
See Supplement 1 for supporting content.

\bibliographystyle{unsrt}

\bibliography{Biblio}

@PREAMBLE{
 "\providecommand{\noopsort}[1]{}" 
 # "\providecommand{\singleletter}[1]{#1}%" 
}

@article{2025persistent,
author = {Qi Yao and Paolo Comaron and Hassan Alnatah and Jonathan Beaumariage and Shouvik Mukherjee and Ken West and Loren Pfeiffer and Kirk Baldwin and Marzena H. Szyma\'{n}ska and David Snoke},
journal = {Optica},
number = {7},
pages = {991--996},
publisher = {Optica Publishing Group},
title = {Persistent, controllable circulation of a polariton ring condensate},
volume = {12},
month = {Jul},
year = {2025},
url = {https://opg.optica.org/optica/abstract.cfm?URI=optica-12-7-991},
doi = {10.1364/OPTICA.560627},
}

@article{comaron2020BKT,
	doi = {10.1209/0295-5075/133/17002},
	url = {https://doi.org/10.1209/0295-5075/133/17002},
	year = {2021},
	month = {jan},
	publisher = {{IOP} Publishing},
	volume = {133},
	number = {1},
	pages = {17002},
	author = {P. Comaron and I. Carusotto and M. H. Szyma{\'{n}}ska and N. P. Proukakis},
	title = {Non-equilibrium Berezinskii-Kosterlitz-Thouless transition in driven-dissipative condensates},
	journal = {Europhysics Letters}
}

@article{Alnatah2024,
	author = {Alnatah, Hassan and Comaron, Paolo and Mukherjee, Shouvik and Beaumariage, Jonathan and Pfeiffer, Loren N. and West, Ken and Baldwin, Kirk and Szyma{\ifmmode\acute{n}\else\'{n}\fi}ska, Marzena and Snoke, David W.},
	title = {{Critical fluctuations in a confined driven-dissipative quantum condensate}},
	journal = {Sci. Adv.},
	volume = {10},
	number = {12},
	year = {2024},
	month = mar,
	issn = {2375-2548},
	publisher = {American Association for the Advancement of Science},
	doi = {10.1126/sciadv.adi6762}
}

@article{Alnatah2024May,
	author = {Alnatah, Hassan and Yao, Qi and Beaumariage, Jonathan and Mukherjee, Shouvik and Tam, Man Chun and Wasilewski, Zbigniew and West, Ken and Baldwin, Kirk and Pfeiffer, Loren N. and Snoke, David W.},
	title = {{Coherence measurements of polaritons in thermal equilibrium reveal a power law for two-dimensional condensates}},
	journal = {Sci. Adv.},
	volume = {10},
	number = {18},
	year = {2024},
	month = may,
	issn = {2375-2548},
	publisher = {American Association for the Advancement of Science},
	doi = {10.1126/sciadv.adk6960}
}

@article{Alnatah2025Jan,
	author = {Alnatah, Hassan and Liang, Shuang and Yao, Qi and Wan, Qiaochu and Beaumariage, Jonathan and West, Ken and Baldwin, Kirk and Pfeiffer, Loren N. and Snoke, David W.},
	title = {{Bose{\textendash}Einstein Condensation of Polaritons at Room Temperature in a GaAs/AlGaAs Structure}},
	journal = {ACS Photonics},
	volume = {12},
	number = {1},
	pages = {48--52},
	year = {2025},
	month = jan,
	publisher = {American Chemical Society},
	doi = {10.1021/acsphotonics.4c01992}
}

@article{Amo2009,
	author = {Amo, Alberto and Lefr{\ifmmode\grave{e}\else\`{e}\fi}re, J{\ifmmode\acute{e}\else\'{e}\fi}r{\ifmmode\hat{o}\else\^{o}\fi}me and Pigeon, Simon and Adrados, Claire and Ciuti, Cristiano and Carusotto, Iacopo and Houdr{\ifmmode\acute{e}\else\'{e}\fi}, Romuald and Giacobino, Elisabeth and Bramati, Alberto},
	title = {{Superfluidity of polaritons in semiconductor microcavities}},
	journal = {Nat. Phys.},
	volume = {5},
	pages = {805--810},
	year = {2009},
	month = nov,
	issn = {1745-2481},
	publisher = {Nature Publishing Group},
	doi = {10.1038/nphys1364}
}

@article{Askitopoulos2013,
	author = {Askitopoulos, A. and Ohadi, H. and Kavokin, A. V. and Hatzopoulos, Z. and Savvidis, P. G. and Lagoudakis, P. G.},
	title = {{Polariton condensation in an optically induced two-dimensional potential}},
	journal = {Phys. Rev. B},
	volume = {88},
	number = {4},
	pages = {041308},
	year = {2013},
	month = jul,
	publisher = {American Physical Society},
	doi = {10.1103/PhysRevB.88.041308}
}

@article{Balili2007,
	author = {Balili, R. and Hartwell, V. and Snoke, D. and Pfeiffer, L. and West, K.},
	title = {{Bose-Einstein Condensation of Microcavity Polaritons in a Trap}},
	journal = {Science},
	volume = {316},
	number = {5827},
	pages = {1007--1010},
	year = {2007},
	month = may,
	issn = {0036-8075},
	publisher = {American Association for the Advancement of Science},
	doi = {10.1126/science.1140990}
}

@article{Beaumariage2024,
	author = {Beaumariage, Jonathan and Sun, Zheng and Alnatah, Hassan and Yao, Qi and Myers, David M. and Steger, Mark and West, Ken and Baldwin, Kirk and Pfeiffer, Loren N. and Tam, Man Chun Alan and Wailewski, Zbig R. and Snoke, David W.},
	title = {{Measurement of exciton fraction of microcavity exciton-polaritons using transfer-matrix modeling}},
	journal = {arXiv},
	year = {2024},
	month = jun,
	eprint = {2406.12940},
	doi = {10.48550/arXiv.2406.12940}
}

@article{Betzold2020,
	author = {Betzold, Simon and Dusel, Marco and Kyriienko, Oleksandr and Dietrich, Christof P. and Klembt, Sebastian and Ohmer, J{\ifmmode\ddot{u}\else\"{u}\fi}rgen and Fischer, Utz and Shelykh, Ivan A. and Schneider, Christian and H{\ifmmode\ddot{o}\else\"{o}\fi}fling, Sven},
	title = {{Coherence and Interaction in Confined Room-Temperature Polariton Condensates with Frenkel Excitons}},
	journal = {ACS Photonics},
	volume = {7},
	number = {2},
	pages = {384--392},
	year = {2020},
	month = feb,
	publisher = {American Chemical Society},
	doi = {10.1021/acsphotonics.9b01300}
}

@article{Bobrovska_2015,
   title={Adiabatic approximation and fluctuations in exciton-polariton condensates},
   volume={92},
   ISSN={1550-235X},
   url={http://dx.doi.org/10.1103/PhysRevB.92.035311},
   DOI={10.1103/physrevb.92.035311},
   number={3},
   journal={Physical Review B},
   publisher={American Physical Society (APS)},
   author={Bobrovska, Nataliya and Matuszewski, Michał},
   year={2015},
   month=jul }

@article{Brune2025,
	author = {Brune, Yannik and Rozas, Elena and West, Ken and Baldwin, Kirk and Pfeiffer, Loren N. and Beaumariage, Jonathan and Alnatah, Hassan and Snoke, David W. and A{\ss}mann, Marc},
	title = {{Quantum coherence of a long-lifetime exciton-polariton condensate}},
	journal = {Commun. Mater.},
	volume = {6},
	number = {123},
	pages = {1--8},
	year = {2025},
	month = jun,
	issn = {2662-4443},
	publisher = {Nature Publishing Group},
	doi = {10.1038/s43246-025-00848-6}
}

@article{Byrnes2014,
	author = {Byrnes, Tim and Kim, Na Young and Yamamoto, Yoshihisa},
	title = {{Exciton{\textendash}polariton condensates}},
	journal = {Nat. Phys.},
	volume = {10},
	pages = {803--813},
	year = {2014},
	month = nov,
	issn = {1745-2481},
	publisher = {Nature Publishing Group},
	doi = {10.1038/nphys3143}
}

@article{Carusotto_2013,
   title={Quantum fluids of light},
   volume={85},
   ISSN={1539-0756},
   url={http://dx.doi.org/10.1103/RevModPhys.85.299},
   DOI={10.1103/revmodphys.85.299},
   number={1},
   journal={Reviews of Modern Physics},
   publisher={American Physical Society (APS)},
   author={Carusotto, Iacopo and Ciuti, Cristiano},
   year={2013},
   month=feb, pages={299–366} }

@article{Choi2022,
	author = {Choi, Daegwang and Park, Min and Oh, Byoung Yong and Kwon, Min-Sik and Park, Suk In and Kang, Sooseok and Song, Jin Dong and Ko, Dogyun and Sun, Meng and Savenko, Ivan G. and Cho, Yong-Hoon and Choi, Hyoungsoon},
	title = {{Observation of a single quantized vortex vanishment in exciton-polariton superfluids}},
	journal = {Phys. Rev. B},
	volume = {105},
	number = {6},
	pages = {L060502},
	year = {2022},
	month = feb,
	publisher = {American Physical Society},
	doi = {10.1103/PhysRevB.105.L060502}
}

@article{Cristofolini2013,
	author = {Cristofolini, P. and Dreismann, A. and Christmann, G. and Franchetti, G. and Berloff, N. G. and Tsotsis, P. and Hatzopoulos, Z. and Savvidis, P. G. and Baumberg, J. J.},
	title = {{Optical Superfluid Phase Transitions and Trapping of Polariton Condensates}},
	journal = {Phys. Rev. Lett.},
	volume = {110},
	number = {18},
	pages = {186403},
	year = {2013},
	month = may,
	publisher = {American Physical Society},
	doi = {10.1103/PhysRevLett.110.186403}
}

@article{Comaron_2025,
   title={Coherence of a non-equilibrium polariton condensate across the interaction-mediated phase transition},
   volume={8},
   ISSN={2399-3650},
   url={http://dx.doi.org/10.1038/s42005-025-01977-7},
   DOI={10.1038/s42005-025-01977-7},
   number={1},
   journal={Communications Physics},
   publisher={Springer Science and Business Media LLC},
   author={Comaron, P. and Estrecho, E. and Wurdack, M. and Pieczarka, M. and Steger, M. and Snoke, D. W. and West, K. and Pfeiffer, L. N. and Truscott, A. G. and Matuszewski, M. and Szymańska, M. H. and Ostrovskaya, E. A.},
   year={2025},
   month=mar }

@article{Dennis_2013,
  title={XMDS2: Fast, scalable simulation of coupled stochastic partial differential equations},
  author={Dennis, Graham R and Hope, Joseph J and Johnsson, Mattias T},
  journal={Computer Physics Communications},
  volume={184},
  number={1},
  pages={201--208},
  year={2013},
  publisher={Elsevier}
}

@article{Dusel2020,
	author = {Dusel, M. and Betzold, S. and Egorov, O. A. and Klembt, S. and Ohmer, J. and Fischer, U. and H{\ifmmode\ddot{o}\else\"{o}\fi}fling, S. and Schneider, C.},
	title = {{Room temperature organic exciton{\textendash}polariton condensate in a lattice}},
	journal = {Nat. Commun.},
	volume = {11},
	number = {2863},
	pages = {2863},
	year = {2020},
	month = jun,
	issn = {2041-1723},
	publisher = {Nature Publishing Group},
	doi = {10.1038/s41467-020-16656-0}
}

@article{Estrecho_2018,
  title={Single-shot condensation of exciton polaritons and the hole burning effect},
  author={Estrecho, E and Gao, Tingge and Bobrovska, Nataliya and Fraser, Michael D and Steger, M and Pfeiffer, L and West, K and Liew, TCH and Matuszewski, Michal and Snoke, David W and others},
  journal={Nature communications},
  volume={9},
  number={1},
  pages={2944},
  year={2018},
  publisher={Nature Publishing Group UK London}
}

@article{Ferrier2011,
	author = {Ferrier, Lydie and Wertz, Esther and Johne, Robert and Solnyshkov, Dmitry D. and Senellart, Pascale and Sagnes, Isabelle and Lema{\ifmmode\hat{\imath}\else\^{\i}\fi}tre, Aristide and Malpuech, Guillaume and Bloch, Jacqueline},
	title = {{Interactions in Confined Polariton Condensates}},
	journal = {Phys. Rev. Lett.},
	volume = {106},
	number = {12},
	pages = {126401},
	year = {2011},
	month = mar,
	publisher = {American Physical Society},
	doi = {10.1103/PhysRevLett.106.126401}
}

@article{Galbiati2012,
	author = {Galbiati, Marta and Ferrier, Lydie and Solnyshkov, Dmitry D. and Tanese, Dimitrii and Wertz, Esther and Amo, Alberto and Abbarchi, Marco and Senellart, Pascale and Sagnes, Isabelle and Lema{\ifmmode\hat{\imath}\else\^{\i}\fi}tre, Aristide and Galopin, Elisabeth and Malpuech, Guillaume and Bloch, Jacqueline},
	title = {{Polariton Condensation in Photonic Molecules}},
	journal = {Phys. Rev. Lett.},
	volume = {108},
	number = {12},
	pages = {126403},
	year = {2012},
	month = mar,
	publisher = {American Physical Society},
	doi = {10.1103/PhysRevLett.108.126403}
}

@article{Guerrero2025,
	author = {Guerrero, Killian and Falque, K{\ifmmode\acute{e}\else\'{e}\fi}vin and Giacobino, Elisabeth and Bramati, Alberto and Jacquet, Maxime J.},
	title = {{Multiply Quantized Vortex Spectroscopy in a Quantum Fluid of Light}},
	journal = {Phys. Rev. Lett.},
	volume = {135},
	number = {24},
	pages = {243801},
	year = {2025},
	month = dec,
	publisher = {American Physical Society},
	doi = {10.1103/whgn-6889}
}

@article{Jin2025,
	author = {Jin, Feng and Ren, Jiahao and Zanotti, Simone and Cui, Peiyuan and Zheng, Hao and Liang, Jie and Ni, Jinyang and Wang, Zhe and Shi, Fulong and Qiu, Cheng-Wei and Chang, Guoqing and Liew, Timothy C. H. and Su, Rui},
	title = {{Exciton polariton condensation in a perovskite moir{\ifmmode\acute{e}\else\'{e}\fi} flat band at room temperature}},
	journal = {Sci. Adv.},
	volume = {11},
	number = {32},
	year = {2025},
	month = aug,
	issn = {2375-2548},
	publisher = {American Association for the Advancement of Science},
	doi = {10.1126/sciadv.adx2361}
}

@article{Kasprzak2006,
	author = {Kasprzak, J. and Richard, M. and Kundermann, S. and Baas, A. and Jeambrun, P. and Keeling, J. M. J. and Marchetti, F. M. and Szyma{\ifmmode\acute{n}\else\'{n}\fi}ska, M. H. and Andr{\ifmmode\acute{e}\else\'{e}\fi}, R. and Staehli, J. L. and Savona, V. and Littlewood, P. B. and Deveaud, B. and Dang, Le Si},
	title = {{Bose{\textendash}Einstein condensation of exciton polaritons}},
	journal = {Nature},
	volume = {443},
	pages = {409--414},
	year = {2006},
	month = sep,
	issn = {1476-4687},
	publisher = {Nature Publishing Group},
	doi = {10.1038/nature05131}
}

@article{Lagoudakis2008,
	author = {Lagoudakis, K. G. and Wouters, M. and Richard, M. and Baas, A. and Carusotto, I. and Andr{\ifmmode\acute{e}\else\'{e}\fi}, R. and Dang, Le Si and Deveaud-Pl{\ifmmode\acute{e}\else\'{e}\fi}dran, B.},
	title = {{Quantized vortices in an exciton{\textendash}polariton condensate}},
	journal = {Nat. Phys.},
	volume = {4},
	pages = {706--710},
	year = {2008},
	month = sep,
	issn = {1745-2481},
	publisher = {Nature Publishing Group},
	doi = {10.1038/nphys1051}
}

@article{Lerario2017,
	author = {Lerario, Giovanni and Fieramosca, Antonio and Barachati, F{\ifmmode\acute{a}\else\'{a}\fi}bio and Ballarini, Dario and Daskalakis, Konstantinos S. and Dominici, Lorenzo and De Giorgi, Milena and Maier, Stefan A. and Gigli, Giuseppe and K{\ifmmode\acute{e}\else\'{e}\fi}na-Cohen, St{\ifmmode\acute{e}\else\'{e}\fi}phane and Sanvitto, Daniele},
	title = {{Room-temperature superfluidity in a polariton condensate}},
	journal = {Nat. Phys.},
	volume = {13},
	pages = {837--841},
	year = {2017},
	month = sep,
	issn = {1745-2481},
	publisher = {Nature Publishing Group},
	doi = {10.1038/nphys4147}
}

@article{Luders2018,
	author = {L{\ifmmode\ddot{u}\else\"{u}\fi}ders, Carolin and Thewes, Johannes and Assmann, Marc},
	title = {{Real time g(2) monitoring with 100 kHz sampling rate}},
	journal = {Opt. Express},
	volume = {26},
	number = {19},
	pages = {24854--24863},
	year = {2018},
	issn = {1094-4087},
	publisher = {Optica Publishing Group},
	doi = {10.1364/OE.26.024854}
}

@article{Luders_2021,
   title={Quantifying Quantum Coherence in Polariton Condensates},
   volume={2},
   ISSN={2691-3399},
   url={http://dx.doi.org/10.1103/PRXQuantum.2.030320},
   DOI={10.1103/prxquantum.2.030320},
   number={3},
   journal={PRX Quantum},
   publisher={American Physical Society (APS)},
   author={Lüders, Carolin and Pukrop, Matthias and Rozas, Elena and Schneider, Christian and Höfling, Sven and Sperling, Jan and Schumacher, Stefan and Aßmann, Marc},
   year={2021},
   month=aug }

@article{McAlister1997,
	author = {McAlister, D. F. and Raymer, M. G.},
	title = {{Ultrafast photon-number correlations from dual-pulse, phase-averaged homodyne detection}},
	journal = {Phys. Rev. A},
	volume = {55},
	number = {3},
	pages = {R1609--R1612},
	year = {1997},
	month = mar,
	publisher = {American Physical Society},
	doi = {10.1103/PhysRevA.55.R1609}
}

@article{Nelsen2013,
	author = {Nelsen, Bryan and Liu, Gangqiang and Steger, Mark and Snoke, David W. and Balili, Ryan and West, Ken and Pfeiffer, Loren},
	title = {{Dissipationless Flow and Sharp Threshold of a Polariton Condensate with Long Lifetime}},
	journal = {Phys. Rev. X},
	volume = {3},
	number = {4},
	pages = {041015},
	year = {2013},
	month = nov,
	publisher = {American Physical Society},
	doi = {10.1103/PhysRevX.3.041015}
}

@article{Olsen_2009,
   title={Numerical representation of quantum states in the positive-P and Wigner representations},
   volume={282},
   ISSN={0030-4018},
   url={http://dx.doi.org/10.1016/j.optcom.2009.06.033},
   DOI={10.1016/j.optcom.2009.06.033},
   number={19},
   journal={Optics Communications},
   publisher={Elsevier BV},
   author={Olsen, M.K. and Bradley, A.S.},
   year={2009},
   month=oct, pages={3924–3929} }

@article{Wouters_2010_1,
   title={Energy relaxation in one-dimensional polariton condensates},
   volume={82},
   ISSN={1550-235X},
   url={http://dx.doi.org/10.1103/PhysRevB.82.245315},
   DOI={10.1103/physrevb.82.245315},
   number={24},
   journal={Physical Review B},
   publisher={American Physical Society (APS)},
   author={Wouters, M. and Liew, T. C. H. and Savona, V.},
   year={2010},
   month=dec }

@article{Wouters_2010_2,
  title = {Superfluidity and Critical Velocities in Nonequilibrium Bose-Einstein Condensates},
  author = {Wouters, Michiel and Carusotto, Iacopo},
  journal = {Phys. Rev. Lett.},
  volume = {105},
  issue = {2},
  pages = {020602},
  numpages = {4},
  year = {2010},
  month = {Jul},
  publisher = {American Physical Society},
  doi = {10.1103/PhysRevLett.105.020602},
  url = {https://link.aps.org/doi/10.1103/PhysRevLett.105.020602}
}

@article{Wouters_2012,
doi = {10.1088/1367-2630/14/7/075020},
url = {https://doi.org/10.1088/1367-2630/14/7/075020},
year = {2012},
month = {jul},
publisher = {IOP Publishing},
volume = {14},
number = {7},
pages = {075020},
author = {Wouters, Michiel},
title = {Energy relaxation in the mean-field description of polariton condensates},
journal = {New Journal of Physics}
}

@article{Gladilin_2020,
   title={Classical field model for arrays of photon condensates},
   volume={101},
   ISSN={2469-9934},
   url={http://dx.doi.org/10.1103/PhysRevA.101.043814},
   DOI={10.1103/physreva.101.043814},
   number={4},
   journal={Physical Review A},
   publisher={American Physical Society (APS)},
   author={Gladilin, Vladimir N. and Wouters, Michiel},
   year={2020},
   month=apr }

@article{Orfanakis2021,
	author = {Orfanakis, K. and Tzortzakakis, A. F. and Petrosyan, D. and Savvidis, P. G. and Ohadi, H.},
	title = {{Ultralong temporal coherence in optically trapped exciton-polariton condensates}},
	journal = {Phys. Rev. B},
	volume = {103},
	number = {23},
	pages = {235313},
	year = {2021},
	month = jun,
	publisher = {American Physical Society},
	doi = {10.1103/PhysRevB.103.235313}
}

@article{Roumpos2013,
	author = {Roumpos, Georgios and Cundiff, Steven T.},
	title = {{Multichannel homodyne detection for quantum optical tomography}},
	journal = {J. Opt. Soc. Am. B, JOSAB},
	volume = {30},
	number = {5},
	pages = {1303--1316},
	year = {2013},
	month = may,
	issn = {1520-8540},
	publisher = {Optica Publishing Group},
	doi = {10.1364/JOSAB.30.001303}
}

@article{Sharma2021,
	author = {Sharma, Vaibhav and Mueller, Erich J.},
	title = {{Driven-dissipative control of cold atoms in tilted optical lattices}},
	journal = {Phys. Rev. A},
	volume = {103},
	number = {4},
	pages = {043322},
	year = {2021},
	month = apr,
	publisher = {American Physical Society},
	doi = {10.1103/PhysRevA.103.043322}
}

@article{Steger2013,
	author = {Steger, Mark and Liu, Gangqiang and Nelsen, Bryan and Gautham, Chitra and Snoke, David W. and Balili, Ryan and Pfeiffer, Loren and West, Ken},
	title = {{Long-range ballistic motion and coherent flow of long-lifetime polaritons}},
	journal = {Phys. Rev. B},
	volume = {88},
	number = {23},
	pages = {235314},
	year = {2013},
	month = dec,
	publisher = {American Physical Society},
	doi = {10.1103/PhysRevB.88.235314}
}

@article{Sun2017,
	author = {Sun, Yongbao and Wen, Patrick and Yoon, Yoseob and Liu, Gangqiang and Steger, Mark and Pfeiffer, Loren N. and West, Ken and Snoke, David W. and Nelson, Keith A.},
	title = {{Bose-Einstein Condensation of Long-Lifetime Polaritons in Thermal Equilibrium}},
	journal = {Phys. Rev. Lett.},
	volume = {118},
	number = {1},
	pages = {016602},
	year = {2017},
	month = jan,
	publisher = {American Physical Society},
	doi = {10.1103/PhysRevLett.118.016602}
}

@article{Sun_2018,
   title={Stable switching among high-order modes in polariton condensates},
   volume={97},
   ISSN={2469-9969},
   url={http://dx.doi.org/10.1103/PhysRevB.97.045303},
   DOI={10.1103/physrevb.97.045303},
   number={4},
   journal={Physical Review B},
   publisher={American Physical Society (APS)},
   author={Sun, Yongbao and Yoon, Yoseob and Khan, Saeed and Ge, Li and Steger, Mark and Pfeiffer, Loren N. and West, Ken and Türeci, Hakan E. and Snoke, David W. and Nelson, Keith A.},
   year={2018},
   month=jan }

@article{Tomita2017,
	author = {Tomita, Takafumi and Nakajima, Shuta and Danshita, Ippei and Takasu, Yosuke and Takahashi, Yoshiro},
	title = {{Observation of the Mott insulator to superfluid crossover of a driven-dissipative Bose-Hubbard system}},
	journal = {Sci. Adv.},
	volume = {3},
	number = {12},
	year = {2017},
	month = dec,
	issn = {2375-2548},
	publisher = {American Association for the Advancement of Science},
	doi = {10.1126/sciadv.1701513}
}

@article{Walker2018,
	author = {Walker, Benjamin T. and Flatten, Lucas C. and Hesten, Henry J. and Mintert, Florian and Hunger, David and Trichet, Aur{\ifmmode\acute{e}\else\'{e}\fi}lien A. P. and Smith, Jason M. and Nyman, Robert A.},
	title = {{Driven-dissipative non-equilibrium Bose{\textendash}Einstein condensation of less than ten photons}},
	journal = {Nat. Phys.},
	volume = {14},
	pages = {1173--1177},
	year = {2018},
	month = dec,
	issn = {1745-2481},
	publisher = {Nature Publishing Group},
	doi = {10.1038/s41567-018-0270-1}
}

@article{Wouters_2009,
   title={Stochastic classical field model for polariton condensates},
   volume={79},
   ISSN={1550-235X},
   url={http://dx.doi.org/10.1103/PhysRevB.79.165302},
   DOI={10.1103/physrevb.79.165302},
   number={16},
   journal={Physical Review B},
   publisher={American Physical Society (APS)},
   author={Wouters, Michiel and Savona, Vincenzo},
   year={2009},
   month=apr }

@misc{mydata,
  author       = { },
  title        = {Replication Data for: "Temporal hopping dynamics in exciton-polariton condensation},
  howpublished = {\url{ https://doi.org/10.17877/TUDODATA-2026-MO8OVXWC}},
  year         = {2026}
}

\end{document}


\title{Temporal hopping dynamics in exciton-polariton condensation: supplemental document}

\author{}

\section{Filtering of mechanical noise}

 The time resolved measurements reveal small amplitude oscillations at fixed frequencies in the acquired quadratures, and thus, the photon number. These oscillations originate from mechanical noise sources in the laboratory, introducing periodic components at well-defined frequencies without influencing the intrinsic condensate dynamics. To remove their contribution, a band stop filter based on a Fourier transform was applied. The filtering process eliminates only the components of the two frequency bands corresponding to the noise peaks, while leaving all other frequencies of the signal unaffected. 

\begin{figure}[h]
            \centering
            \includegraphics[width=0.4\columnwidth]{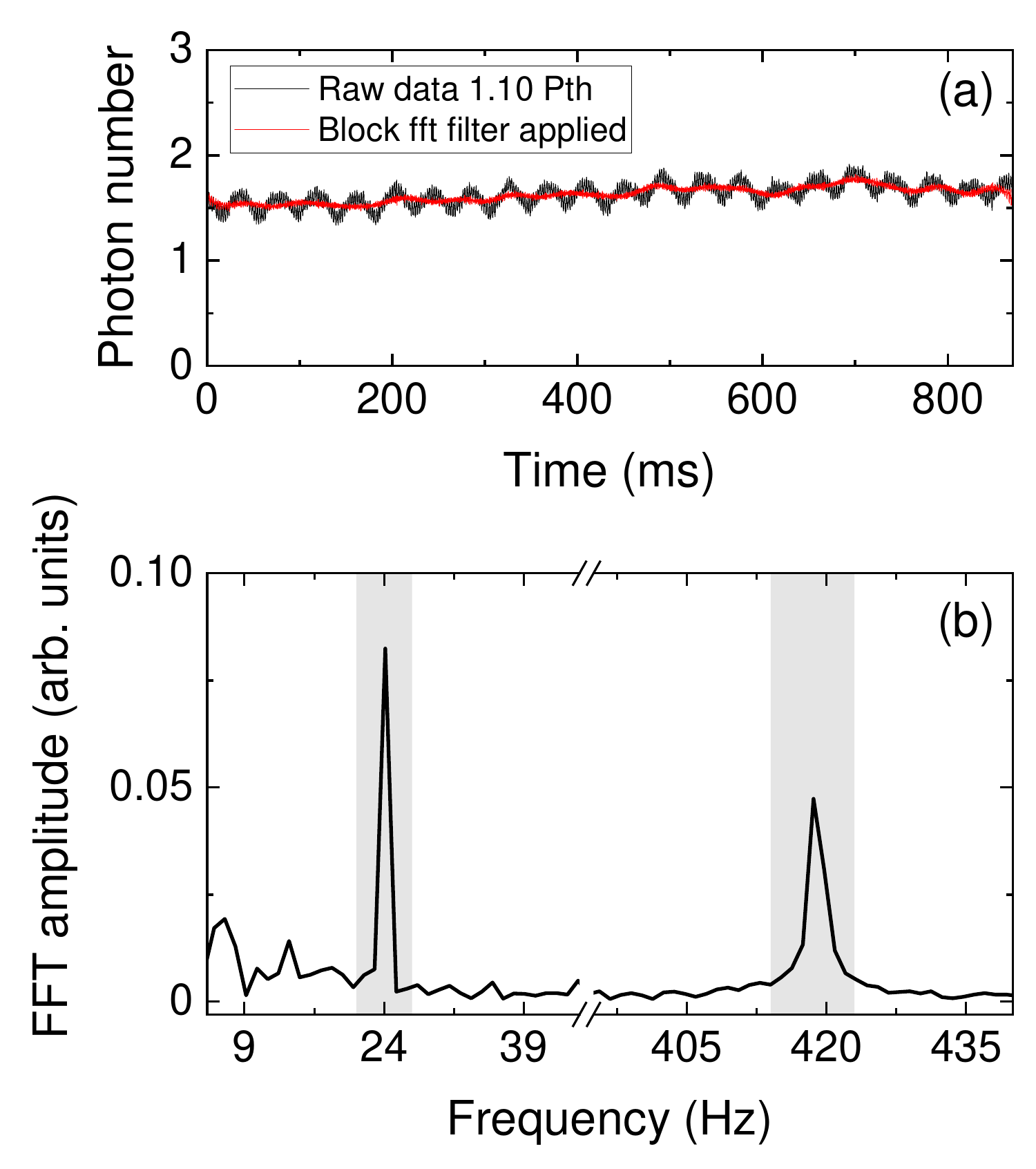}
            \caption{\textbf{Removal of mechanical noise.} (a) Time-resolved photon number at $1.10\, P_{th}$ before (black line) and after (red line) the filtering process. (b) Fourier transform amplitude of the unfiltered data shown in (a), where two pronounced peaks at 24 Hz and 418 Hz are identified as mechanical noise contributions. The shaded areas indicate the frequency bands removed by the FFT filter.}
            \label{fig:FigS1}
        \end{figure}

 Figure \ref{fig:FigS1} illustrates this procedure. Panel (a) shows the unprocessed photon-number (black line) recorded at $1.10\, P_{th}$, where periodic oscillations are clearly visible. The Fourier transform of this signal, shown in panel (b), reveals two dominant peaks centered at 24 Hz and 418 Hz. Then, a band filter to block these frequencies and the shaded areas around them is applied. The resulting photon-number is recalculated and displayed in panel (a) (red line), showing that the mechanical oscillations are effectively suppressed without altering the underlying signal behavior. Note that the calculation of $g^{(2)}(0)$ employs a bin size much shorter than the oscillation period of these noise components. Therefore, the filtering process does not affect the correlation function.

\section{Satisfying Truncated Wigner Approximation Constraints}

As detailed in the main text, the characterization of thermal and condensed states via the Truncated Wigner Approximation (TWA) observable $g^{(2)}(0)$ requires the polariton field to avoid the pathological density $1/2dV$. Moreover, it is important to note that the TWA formalism is most reliable in a highly occupied regime, particularly at stronger pumping powers. Together, these considerations introduce a numerical constraint that influences several aspects of the modeling and analysis.
One consequence is the use of the nonrenormalized density in the Truncated Wigner equation, which increases the average densities by approximately $10\%$. The qualitative dynamics and the modes involved remain unchanged, but the resulting higher densities place the analysis further away from the influence of the problematic $1/2dV$.
The decision to use a substantially smaller excitonic fraction, with $|X|^{2} = 0.1$ being employed instead of the experimental $|X|^{2} \approx0.52$, is another adjustment used to ensure higher densities. Since polariton-polariton interactions arise from their excitonic component, the interaction induced blueshift, given by $g_{\mathrm{c}}|\psi|^2$, strongly depends on the value of $|X|^{2}$. Its reduction weakens the effective nonlinearity, which then calls for a larger occupation to achieve gain saturation \cite{Carusotto_2013}.

\begin{figure}[h]
    \centering
    \includegraphics[width=0.75\columnwidth]{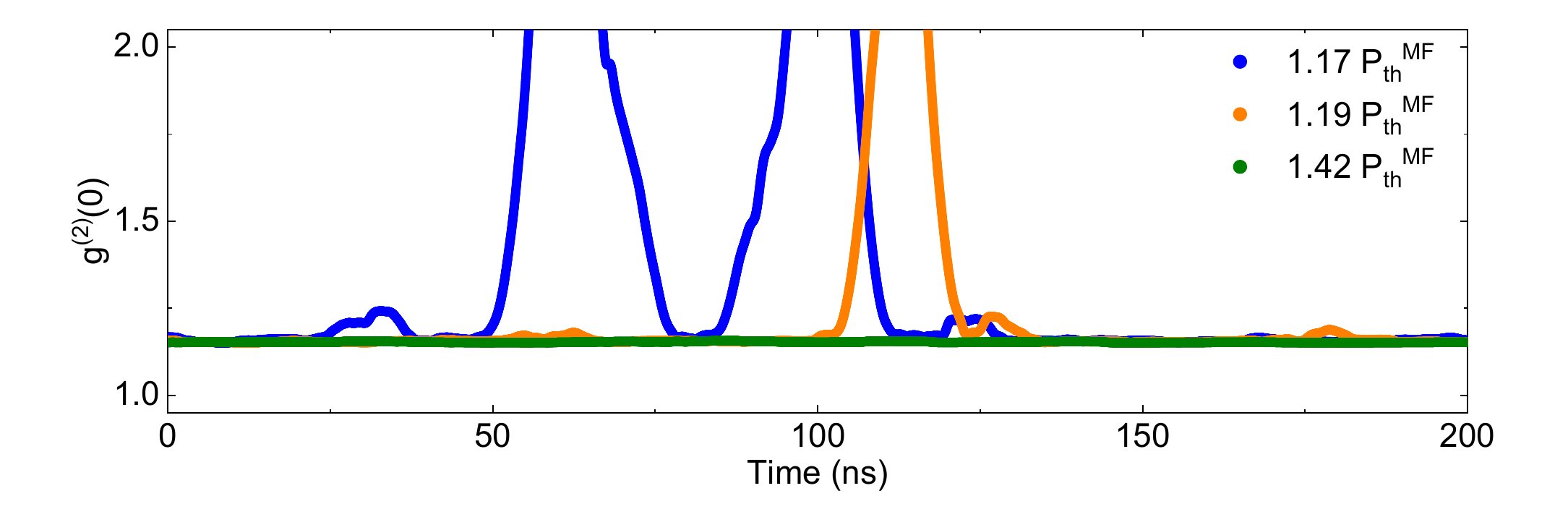}
    \caption{\textbf{Evaluation of the uncorrected $g^{(2)}(0)$ observable at probing points $1.17 \, P_{th}^{\mathrm{MF}}$, $1.19 \, P_{th}^{\mathrm{MF}}$, and $1.42 \, P_{th}^{\mathrm{MF}}$} These results closely match those obtained using the corrected expression presented in Fig. 3 of the main text. For cases $1.17 \, P_{th}^{\mathrm{MF}}$ and $1.19 \, P_{th}^{\mathrm{MF}}$, the jump remains finite, with the thermal state values reaching a peak at $g^{(2)}(0) \approx 4$.}
    \label{fig:FigS2}
\end{figure}

\begin{figure}[h]
    \centering
    \includegraphics[width=0.75\columnwidth]{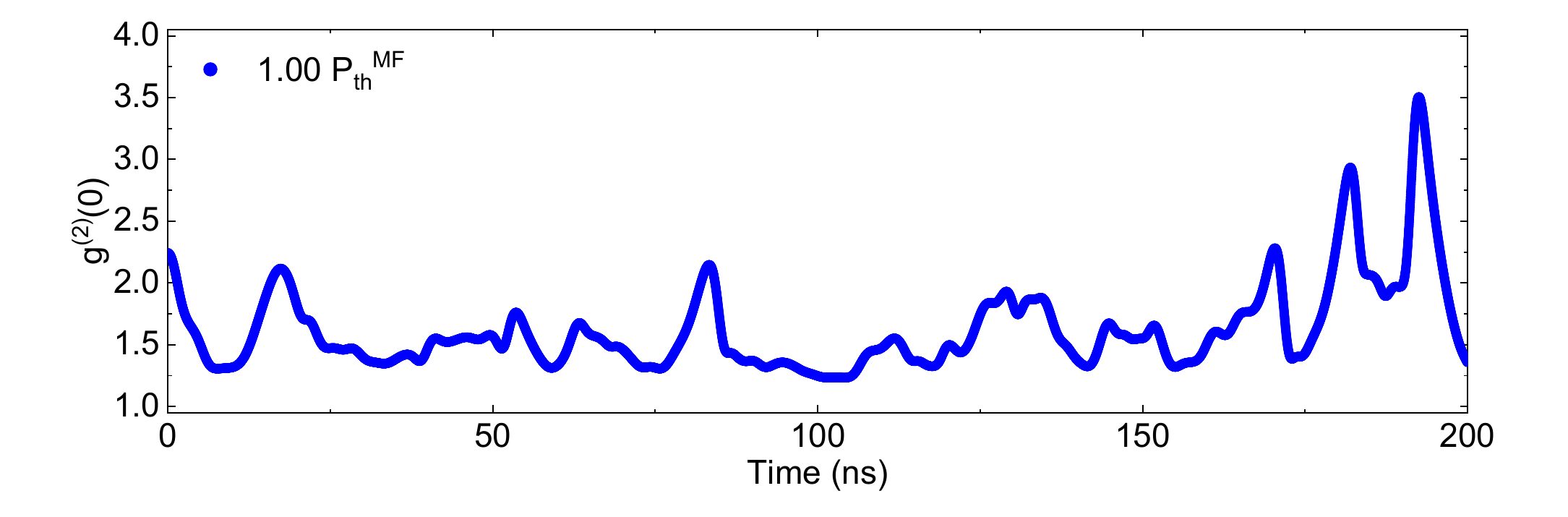}
    \caption{\textbf{Evaluation of the uncorrected $g^{(2)}(0)$ observable for probing point $1.19 \, P_{th}^{\mathrm{MF}}$} The behavior deviates significantly from the expectation for a thermal state, demonstrating the vital role of commutator corrections in recovering physically meaningful results at low occupations.}
    \label{fig:FigS3}
\end{figure}

If the analysis is restricted to the high occupation regime, where the condensate settles and hopping subsides, the influence of the Wigner commutator corrections diminishes even for the $g^{(2)}(0)$ observable. Fig. \ref{fig:FigS2} presents the results obtained using the uncorrected expression $g^{(2)}(0) = \langle \psi^*\psi^*\psi\psi \rangle_W / \langle\psi^*\psi\rangle_{W}^{2}$ for cases of high occupation. The behavior closely follows that of the corresponding points $1.17 \, P_{th}^{\mathrm{MF}}$, $1.19 \, P_{th}^{\mathrm{MF}}$, and $1.42 \, P_{th}^{\mathrm{MF}}$ in Fig. 3 of the main text, with increasing agreement as pumping power is increased. Here, the superscript MF refers to the mean field pumping threshold, obtained through Eqs. (2-3) of the main text without the inclusion of the stochastic noise term dWc. Notably, this form of the observable does not suffer from the divergence and asymptoticity problems introduced by the corrections. Their inclusion in the main analysis is thus motivated by the desire to characterize the negligible occupation regime, for which the corrections fully restore expected behavior. This is demonstrated by Fig. \ref{fig:FigS3}, where the uncorrected observable fails to reproduce the $g^{(2)}(0) \approx 2$ expected for a thermal state.

The filtering procedures, namely the frequency space mode selection, the removal of poorly behaved densities, and the use of a spatial mask, are all designed to prevent the contamination and biasing of operator averages towards problematic values. In practice, these procedures primarily restrict the analysis to the well behaved and reliable high occupation regime, whilst simultaneously revealing the fully thermal regime associated with negligible densities, both of which are well characterized by the TWA $g^{(2)}(0)$ expression.

\section{Difference Between Simulation Parameters and Experimental Configuration}

Besides the reduced $|X|^{2}$, the other important difference concerns the circumference of the pumping ring, which is taken to be approximately six times larger than in the experiment. This enlargement is imposed by computational constraints arising from the use of a fourth-order Runge–Kutta (RK4) scheme for explicit time evolution in XMDS2 \cite{Dennis_2013}. 

For the RK4 integration of the kinetic term in Eq. (2) of the main text, the discretized laplacian imposes a Von Neumann stability condition  of the form $\Delta t < \Delta x^2 / 2C$, where $\Delta t$ and $\Delta x$ denote the physical time step and lattice spacing, respectively, and where $C = \hslash/2m$. Adimensionalisation only serves to rescale the condition through the removal of $C$; for parameters typically employed in polariton simulations, C is already $\mathcal{O}(1)$, so the stability constraint is not appreciably relaxed.

Accurate modeling of a smaller ring requires reducing $\Delta x$ in order to properly resolve the pump profile. The stability condition then forces a corresponding quadratic reduction of $\Delta t$. Consequently, the number of required time steps increases as $1/\Delta x^{2}$, substantially extending the total computational time, either through longer individual simulations or the need to concatenate multiple shorter runs. 

The larger pumping ring therefore represents a numerical compromise by preserving experimentally relevant physical parameters while keeping the simulation computationally tractable.

\section{Numerical Time Integration and Window Sizing}

As mentioned in the main text, the resolution of hopping events, and thus density fluctuations, depends strongly on the window size applied during the temporal averaging. For the evaluation of the $g^{(2)}(0)$ observable, a window size of 500 samples was used to compute a moving mean. This choice represents a compromise between averaging over a sufficient number of samples to ensure convergence, and maintaining a window small enough to resolve sustained hopping events.

The same long-lived switching events must also be captured by the image difference observable, and accordingly the same window size was adopted. In this case, however, the time series is partitioned into non-overlapping windows of 500 samples, each of which is averaged to produce a single image. The resulting image difference values are then further averaged over five stochastic realizations at each pumping power to improve statistical convergence.

\bibliographystyle{unsrt}

\bibliography{Biblio}